\documentclass[twoside,twocolumn]{article}
\usepackage{blindtext} 
\usepackage{graphicx,txfonts}
\usepackage[sc]{mathpazo} 
\usepackage[T1]{fontenc} 
\usepackage{microtype} 

\usepackage[english]{babel} 

\usepackage[hmarginratio=1:1,top=32mm,columnsep=20pt]{geometry} 
\usepackage[hang, small,labelfont=bf,up,textfont=it,up]{caption} 
\usepackage{booktabs} 

\usepackage{lettrine} 

\usepackage{enumitem} 
\setlist[itemize]{noitemsep} 

\usepackage{abstract} 

\usepackage{titlesec} 
\renewcommand\thesection{\Roman{section}} 
\renewcommand\thesubsection{\roman{subsection}} 
\titleformat{\section}[block]{\large\scshape\centering}{\thesection.}{1em}{} 
\titleformat{\subsection}[block]{\large}{\thesubsection.}{1em}{} 

\usepackage{fancyhdr} 
\usepackage{titling} 

\title{{\huge Mining Artifacts in Mycelium SEM Micrographs}} 
\author{%
\textsc{Thaicia Stona de Almeida} \\[1ex] 
\normalsize University at Buffalo, SUNY \\ 
}
\date{}
\renewcommand{\maketitlehookd}{%
\begin{abstract}
\noindent Mycelium is a promising biomaterial based on fungal mycelium, a highly porous, nanofibrous structure. Scanning electron micrographs are used to characterize its network, but the currently available tools for nanofibrous microstructures do not contemplate the particularities of biomaterials. The adoption of a software for artificial nanofibrous in mycelium characterization adds the uncertainty of imaging artifact formation to the analysis. The reported work combines supervised and unsupervised machine learning methods to automate the identification of artifacts in the mapped pores of mycelium microstructure.\\\\ Keywords: Machine learning; unsupervised learning; image processing; mycelium; microstructure informatics
\end{abstract}
}

\begin{document}
\maketitle
\section{Introduction}
Mycelium is a renewable biomaterial, a composite considered an alternative to synthetic polymers. Mushroom mycelium is the root structure of mushrooms. It consists of a porous structure of nanofibers, the hyphae, which have typical length in the range of a few microns to several meters, depending on the species and growth conditions [6]. The initial growth of the mycelium is isotropic, starting from a spore. After the initial phase, a branching leads to fractal tree-like structure, and fibers randomly start to connect through hyphal fusion [6], creating a heterogeneous network structure. Mechanical properties of the mycelium network are determined by individual hyphae behavior and the topological arrangement of the fibers within the network. Density, Young modulus and yield strength of the mycelium have been shown [7] to correspond to the mechanical behavior of the group of cellular solids, in particular, the open cell foam.

Mycelium can be prepared by growing the structure in molds: agricultural waste, nutrients and liquid mushroom mycelium are mixed and put in a mold. Once its growth has achieved the desired size, the material is demolded and baked. The baking process makes the material inert and dry, killing the mushroom and keeping the designed mold shape. When disposed, it can be used as plant nutrient. Mycelium has been employed in faux leather, structural boards and packaging [11].

\begin{figure}
    \centering
    \includegraphics[scale=0.11]{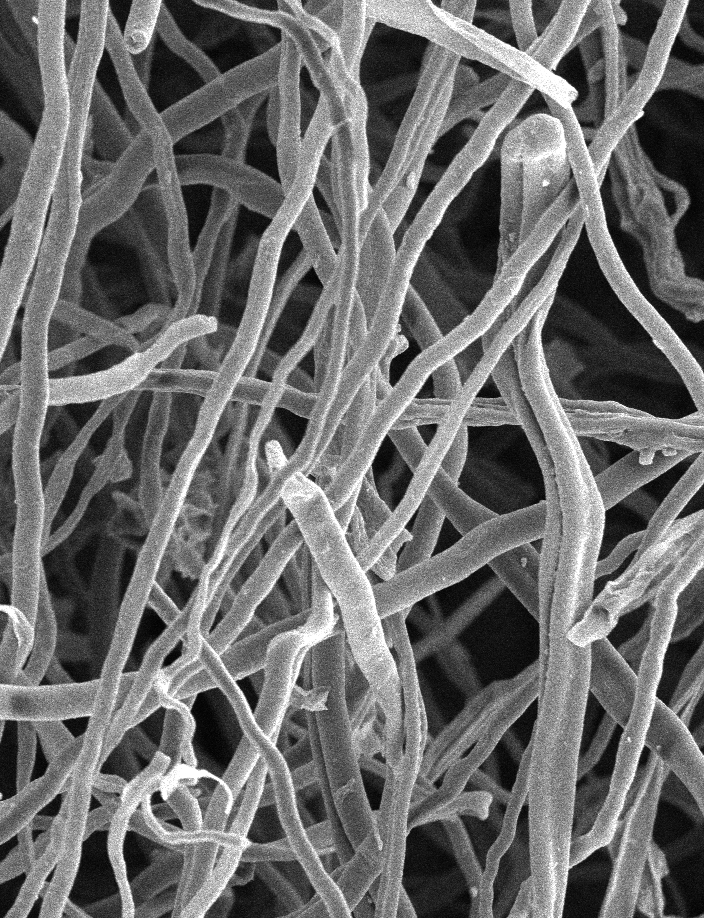}
    \includegraphics[scale=0.11]{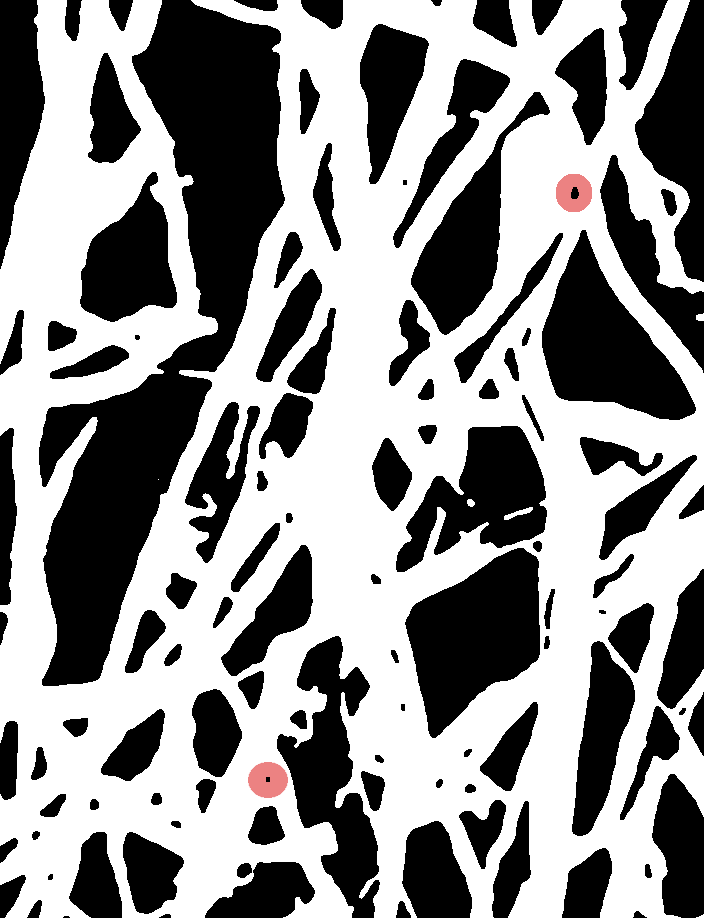}
    \caption{Detail of original micrograph (3000X mag.) and its respective segmentation, with generated artifact pores highlighted}
\end{figure}
\begin{figure*}
    \centering
    \includegraphics[trim=0 0 0 20,clip,width=0.8\textwidth]{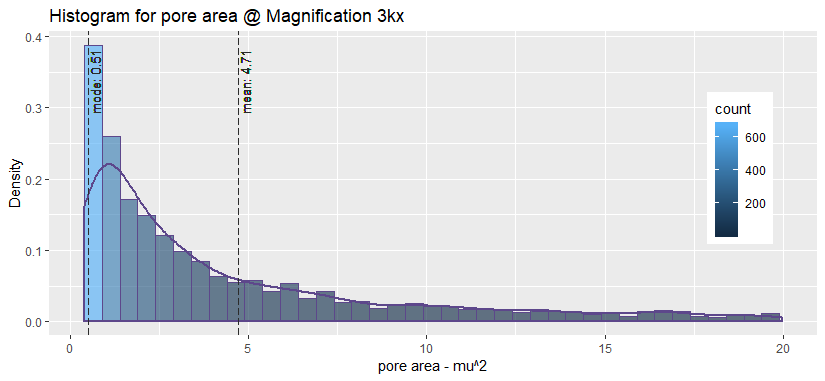}
    \caption{Histogram of pore area after lower outlier cutoff}
\end{figure*}

Porous filtration membranes have wide applications and the fibrous structure of the mycelium potentially can be adopted as a filtration membrane for a natural, biodegradable membrane filter. The permeability of fibrous type of materials have been studied before (ref. [4,5,9]) though the sensitivity to the complexity of the geometry remains a challenge and no general law for permeability has been established. In order to understand the feasibility of such application, it is necessary to characterize the structure of the mycelium in terms of its porosity, permeability, tortuosity and further properties. With this goal in mind, SEM micrographs of \textit{Ganoderma resinaceum} mycelium samples were processed with ImageJ and DiameterJ [2]. DiameterJ is a validated nanofiber characterization tool designed for artificial microstructures (electrospun fibers), and was adopted due to the scarcity of more appropriate materials informatics software for biological nanofibrous microstructure characterization. This work presents a protocol for identifying the most common artifact found in our results, imaging artifacts generated by the segmentation method adopted in the DiameterJ image analysis protocol.

\section{Quality of image analysis for structural characterization}

The choice of segmentation filters and methods for image analysis is typically done by expert. A "good" choice of filter is susceptible to both human and machine-originated uncertainty, for this reason methods for automated and highthroughput characterization of heterogeneous structure such as mycelium are challenging. Specifically, irregular fibers that are not perfectly tubular or smooth, fiber overlapping and intersections added to the light artifacts of SEM microscopy, contribute to the difficulty of the characterization from SEM micrographs.

Even if the accuracy of a method by the human expert is considered satisfactory at higher magnification, its application to lower magnifications can increase the perception of solid material - fibers - corrupting estimated values for fiber width and tortuosity, affecting the pore quality as consequence. On the other hand, filters that can better identify fibers at lower resolutions will necessarily eliminate background fibers, since at lower magnifications more fibers at different depths can be seen under the same focus, and so the solution is to find better contrast between fibers that present similar gray values. Such filter would therefore partially erase the most external parts of the fibers, producing thinner fibers. Moreover, folds and intersection on fibers can affect the extraction of topological features: once the SEM micrograph is segmented into two-phases (fibers and voids) the depth information is lost. \textbf{Fig. 1}: segmented background represents pores (black) and the white phase represents the fiber network. The images analyzed were prepared with magnification 3000X.

The systematic error observed for non-trained DiameterJ operators has been considerably reduced after completion of the DiameterJ online training [8], showing that the human factor can greatly affect the quality of the characterization. This fact suggests an opportunity for machine learning improvement of the segmentation choice process. In order to avoid adding to systematic error, no hand edition was adopted to improve the quality of the segmented images.

\section{Methods}

\subsection{Image processing protocol}

Image processing with DiameterJ is executed in two steps: segmentation with \textit{plugin $\rightarrow$ DiameterJ $\rightarrow$ DiameterJ Segment}, and characterization with \textit{plugin $\rightarrow$ DiameterJ $\rightarrow$ DiameterJ 1-018}, where the best segmented files to be analyzed are chosen by the operator. Open ImageJ and set the scale for one image (SEM micrograph) that belongs to the folder to be analyzed. The scale will be used in the second step of DiameterJ process, automatically converting pixel into micrometer. The step of characterization includes skeletonization and extraction of statistics from the selected segmented images with \textit{DiameterJ 1-018}. The image folder can be batch processed, in which case the output is a set of measurements, statistics and processed images. Once the steps are completed, all the features listed in [2] are available on folder \textit{Best Segmented}. Pore area is the area of the pore determined by mycelium section analyzed, i.e., it is a 2-dimensional measure of pore size. The pore area computed by DiameterJ considers the number of pixels in a black pixel cluster, and converts it to micrometer according to the scale set before when the protocol is initialized. The pore data extracted was considered and a minimum, reasonable value of $0.4\mu m^2$ for pore area was set for lower outlier removal, based on previous expert knowledge. DiameterJ identifies a pore for every contiguous 10-12 or more black pixels [2]. For magnification 3000X, the minimum area detected is of approximately $0.02\mu m^2$, 20 times smaller the adopted lower cutoff. Lower outliers can also be artifacts (the minimum contiguous area detected by DiameterJ is 10px) and not relevant for the proposed future application (filtering). No artifact upper outliers were detected for pore area or radius. The interval considered therefore contains real and artifact pores, with area $\geq 0.4\mu m^2$ (\textbf{Fig. 2}). The consistency of sample statistics observed suggests that samples at considered magnification 3000X can be considered as the representative volume element (RVE) for the analysis.

\subsection{Mining artifacts}

Supervised inspection of one micrograph, its segmented image, and pore map revealed 20 artifact pores (\textbf{Fig. 9}) out of 209 pores, divided in 2 classes: shade pores, corresponding to a shade caused by fiber irregularity or overlap, which creates a blurred dark area, and overlap pore, which correspond to an actual void space between two almost parallel fibers, creating a long, thin sharp dark area.

\begin{figure}[ht]
    \includegraphics[trim=20 0 0 23,clip, scale=0.5]{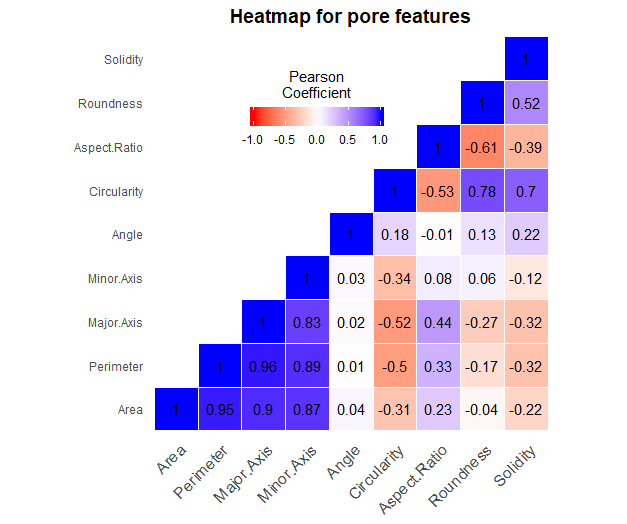}
    \caption{Heatmap of extracted pore features}
\end{figure}

A dataframe containing extracted features of the 209 pores plus the classified artifact pores (\textit{real}, \textit{shade} or \textit{overlap}) was prepared and correlations between quantitative pore features - area, perimeter, major and minor axis, angle, circularity, aspect ratio, roundness and solidity - were assessed with a heatmap (Euclidian distance clustered correlogram).

\begin{figure}[ht]
    \includegraphics[trim=23 0 95 38,clip,scale=0.45]{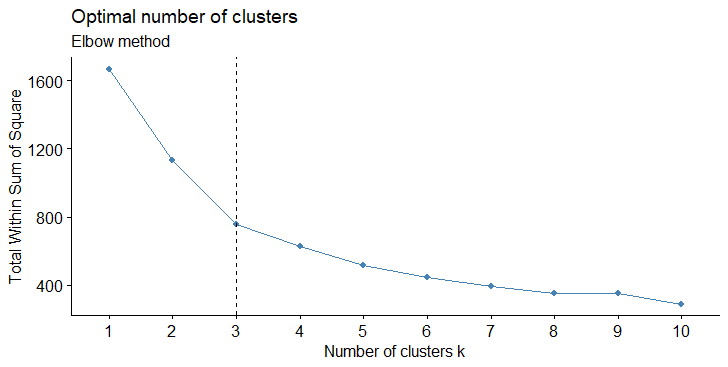}
    \caption{k selection based on total within sum of square}
\end{figure}

The heatmap (\textbf{Fig. 3}) suggests that not all features are necessary to represent the whole dataset of features: the first block is highly correlated - area, perimeter, major and minor axis - angle is an independent feature, the second block is moderately correlated - circularity, AR and roundness - and solidity is weakly correlated to the second block.

Cluster analysis was then employed to study the relation between the classified pores and the features. The feature angle (axis orientation of pores) was removed from the dataset because of its uniform and independent character, and the new dataset was tested for optimal number of clusters using kmeans clustering method. The optimal number of clusters identified for the new 8-dimensional dataset is $k=3$ (\textbf{Fig. 4}). The clustered dataset is presented in the PCA plot of 2-d principal components space (\textbf{Fig. 5}).

\section{Results and Discussion}

The tendency of the features to depend on two main blocks (\textbf{Fig. 3}) is confirmed by the cluster plot, where the PC1 and PC2 represent 82.8$\%$ of the dataset dispersion. Two features of each dominant block of the heatmap were chosen to get a deeper understanding of the cluster characteristics: log of pore area (block 1) and roundness (block 2) (\textbf{Fig. 6}).  Cluster 3 is clearly composed of smaller pores, and is relatively independent of roundness. The cluster 2 is the smallest cluster, containing only 8 points. It is composed of the biggest area pores, and is independent of roundness.

\begin{figure}[ht]
    \centering
    \includegraphics[trim=7 5 5 20,clip,width=0.54\textwidth]{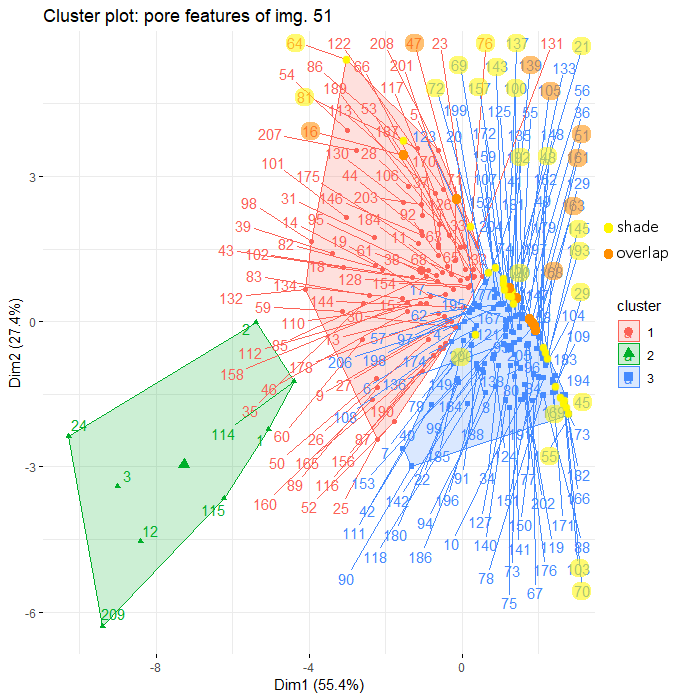}
    \caption{PC plot of clusters}
\end{figure}

The cluster 1 contains the most asymmetric pores, presenting a wide but plausible range of pore area. The density of cluster 1 pore area (\textbf{Fig. 7}) shows a behaviour which is similar to the densities of data curated by an expert with cutoffs (\textbf{Fig. 8}). In fact, cluster 1 contains a much smaller quantity of very small pores. Thus, the observations agree with the discussed arguments in the image processing protocol, presenting a different and method to select representative pores.

The majority of the pores classified as artifacts is concentrated in cluster 3: 25 out of the 30 artifacts, of both classes, shade and overlap, belong to cluster 3. This indicates that setting appropriate lower cut limits for the pore area functions as a practice of data cleansing, reducing the number of artifacts. The fact that no artifacts belong to cluster 2 is consistent with the assumptions, since bigger pores have been found to be accurately identified by DiameterJ. 

\begin{figure}[ht]
    \centering
    \includegraphics[trim= 20 0 0 20 ,clip,scale=0.37]{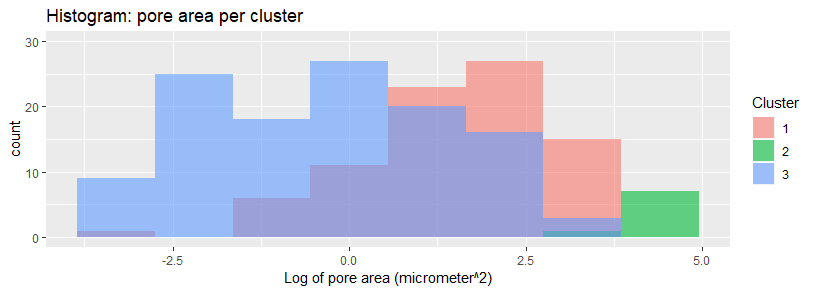}
    \includegraphics[trim= 20 0 0 20 ,clip,scale=0.37]{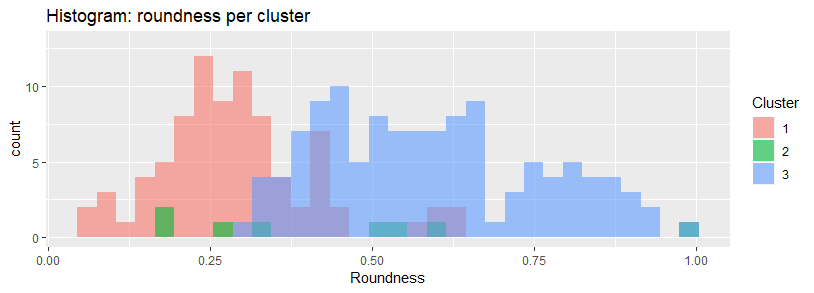}
    \caption{Histograms of log of pore area and pore roundness per cluster}
\end{figure}

 Both classes of artifacts were found on clusters 1 and 3, suggesting that the features presented are insufficient to predict or explain the artifact classes.

\begin{figure*}
    \centering
    \includegraphics[trim= 20 0 0 30 ,clip,width=\textwidth]{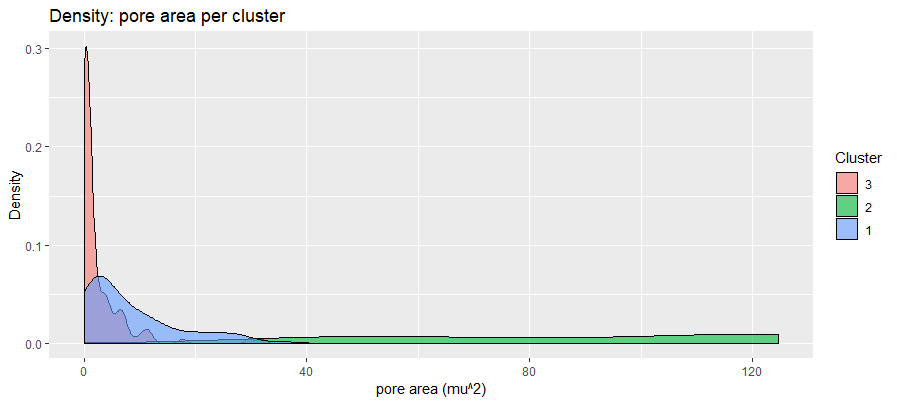}\\
    \caption{Densities of pore area per cluster}
    \centering
    \includegraphics[trim= 20 0 0 20 ,clip,width=0.75\textwidth]{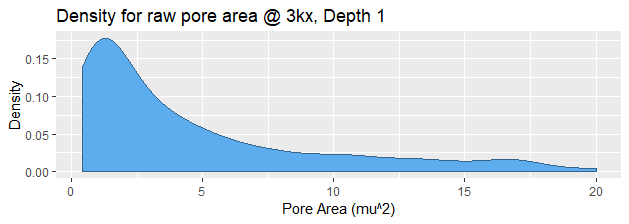}
    \caption{Density of raw pore area for all images at magnification 3kx, values greater than $0.4\mu m^2$}    
\end{figure*}

\begin{figure*}
    \centering
    \includegraphics[width=0.8\textwidth]{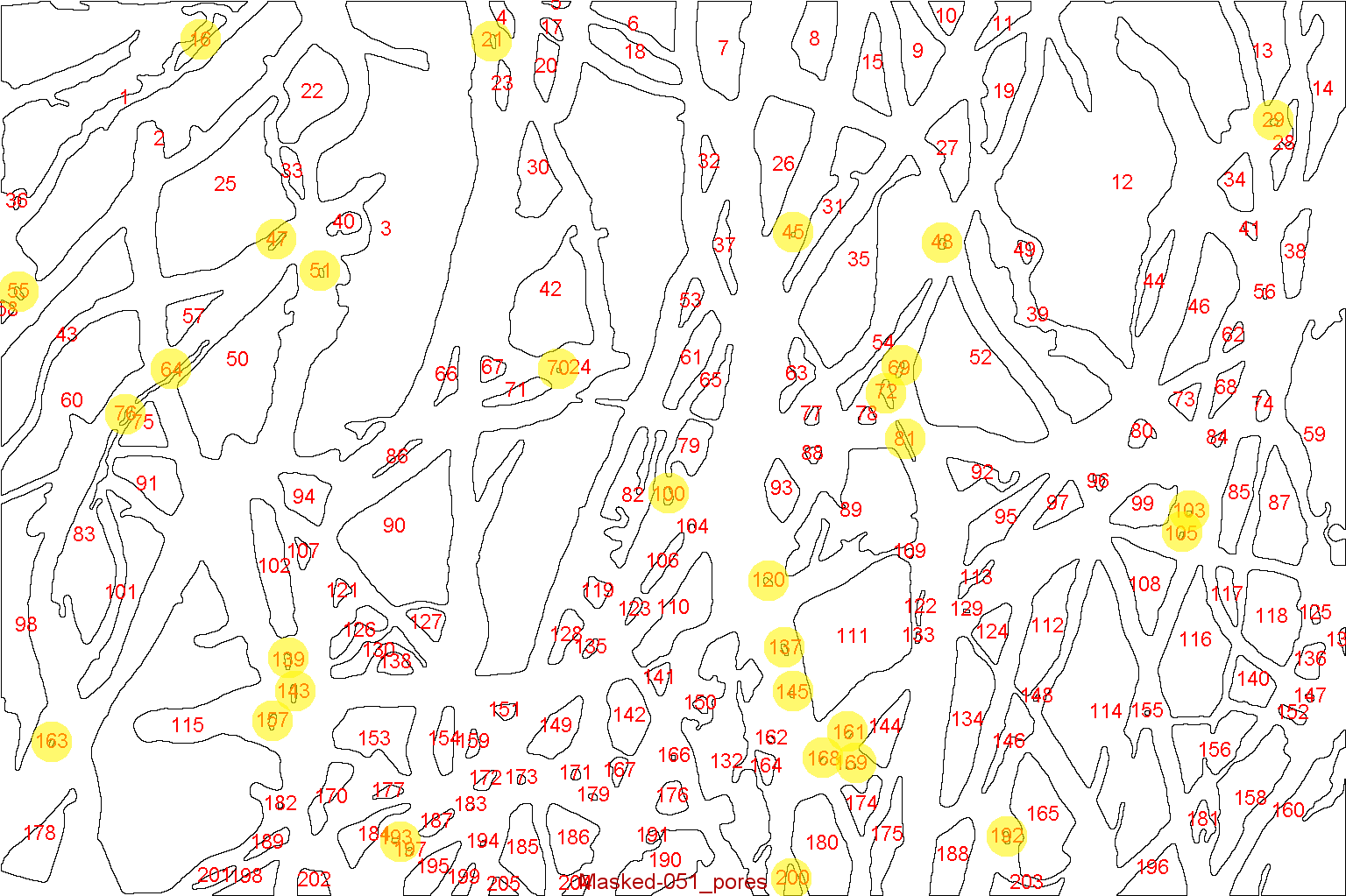}
    \caption{Annotated pore map}
\end{figure*}

It should be noted that the results for artifact pore are limited to the statistics extracted from one SEM image only, and not the whole dataset of micrographs analyzed in the study.

Other limitations of the protocol carried out are the identification of pores by a sole trained specialist, which can be assigned to a team of several and more experienced trained specialists, with an additional step to double check the identified artifacts. More images are to be analyzed. Other classes of artifacts can be defined and the presented classes can be redefined; these could account for the human systematic error factor.

Besides the human factor, and considering that the supervised classification was performed on the original SEM image, while the clustering was performed on the segmented image - thus the loss of information is implicit - since simple classification (Naive Bayes Classification) would require stronger assumptions and a bigger dataset, the method chosen was k-means. K-means was chosen for its simplicity, stability and for the fact that all features analyzed are continuous real numbers, and no additional assumptions were made, which avoids increase of systematic error. The classes of pores were not included in the dataset clustered to avoid the need of the Gower metric, which can aggregate quantitative and qualitative data, but costs additional systematic error. Best results are expected for the application of Naive Bayes Classification run on a larger dataset, classified by several experienced trained specialists.

\section{Conclusion}

The proposed protocol for automated detection of imaging artifacts is consistent with the analytical approach of data cleaning based on expert knowledge and its extension to a larger dataset and double-supervised classification of artifacts by experts is encouraged. Transfer learning could be a potential machine learning method to generalize and automate the classification to future micrograph additions to the database.

\section{Acknowledgements}

TSA thanks P. Nalam and O. Wodo for the helpful discussions and materials provided.

\section{Statement}

There are no competing interests to be declared.



 



\begin{thebibliography}{9}


\bibitem{1}
  Zhao C, Zhou X, Yue Y,
  \emph{Determination of pore sizer and pore size distribution on the surface of hollow-fiber filtration membranes: a review of methods}.
  Desalination 129 (2000) 107-123


\bibitem{2}
    Hotaling NA, Bharti K, Kriel H, Simon Jr CG \emph{DiameterJ: A validated open source nanofiber diameter measurement tool.} Biomaterials 61 (2015) 327–38


\bibitem{3}
  Knackstedt MA, Arns CH, Saadatfar M, Senden TJ, Limaye A, Sakellariou A, Sheppard AP, Sok RM, Schrof W, Steininger H,
  \emph{Elastic and transport properties of cellular solids derived from three-dimensional tomographic images}.
  Proceedings of the Royal Society A, 462 (2006) 2833-2862


\bibitem{4}
  Vallabh R, Banks-Lee P, Seyam A-F,
  \emph{New Approach for Determining Tortuosity in Fibrous Porous Media}.
  Journal of engineered fibers and fabrics 5(3) (2010) 7-15


\bibitem{5}
  Scheidegger AE,
  \emph{The Physics of Flow Through Porous Media}.
  University of Toronto Press (1960)


\bibitem{6}
  Haneef M, Ceseracciu L, Canale C, Bayer IS, Guerrero JAH, Athanassiou A,
  \emph{Advanced Materials From Fungal Mycelium: Fabrication and Tuning of Physical Properties}.
  Scientific Reports 7:41292 January (2017)


\bibitem{7}
  Islam MR, Tudryn G, Bucinell R, Schadler L, Picu RC,
  \emph{Morphology and mechanics of fungal mycelium}.
  Scientific Reports 7:13070 October (2017)


\bibitem{8}
    Hotaling NA, Jeon J, Wade MB, Luong D, Palmer X-L, Bharti K, Simon CGJr,
    \emph{Training to Improve Precision and Accuracy in the Measurement of Fiber Morphology}.
    PLoS ONE 11(12) (2016)


\bibitem{9}
    Koponen A, Kataja M, Timonen J, Kandhai D,
    \emph{Simulations of Single-Fluid Flow in Porous Media},
    International Journal of Modern Physics C,  9 - 8 (1998) 1505-1521


\bibitem{10}
    Rezakhaniha R, Agianniotis A, Schrauwen JT, Griffa A, Sage D, Bouten CV, van de Vosse FN, Unser M, Stergiopulos N,
    \emph{Experimental investigation of collagen waviness and orientation in the arterial adventitia using confocal laser scanning microscopy},
    Biomechanics and modeling in mechanobiology,  11 (2012) 461-473

\bibitem{11}
    Jones M, Huynh T, Dekiwadia C, Daver F, John S,
    \emph{Mycelium Composites: A Review of Engineering Characteristics and Growth Kinetics},
    Journal of Bionanoscience,  11 (2017) 241-257 
    
\bibitem{12}
    Zhao X, Toksoz MN,
    \emph{Modeling Fluid Flow In Heterogeneous And Anisotropic Porous Media},
     Earth Resources Laboratory Industry Consortia Annual Report, 11 (1991)

\end{thebibliography}
\end{document}